\documentclass{Odyssey2026}

\usepackage{amsmath,graphicx,subcaption}
\usepackage{enumitem}
\usepackage[disable]{todonotes}
\usepackage{booktabs}
\usepackage{placeins}

\usepackage{booktabs}
\usepackage{multirow}
\usepackage{colortbl}
\usepackage[table]{xcolor}

\definecolor{auc5}{RGB}{142,206,179}  %
\definecolor{auc4}{RGB}{195,228,195}  %
\definecolor{auc3}{RGB}{197,218,238}  %
\definecolor{auc2}{RGB}{245,220,170}  %
\definecolor{auc1}{RGB}{237,183,183}  %
 
\newcommand{\C}[1]{%
  \ifdim #1 pt > 0.849 pt \cellcolor{auc5}\color[RGB]{8,80,65}#1%
  \else\ifdim #1 pt > 0.749 pt \cellcolor{auc4}\color[RGB]{8,80,65}#1%
  \else\ifdim #1 pt > 0.649 pt \cellcolor{auc3}\color[RGB]{8,80,65}#1%
  \else\ifdim #1 pt > 0.549 pt \cellcolor{auc2}\color[RGB]{8,80,65}#1%
  \else \cellcolor{auc1}\color[RGB]{8,80,65}#1%
  \fi\fi\fi\fi}

\newcommand{\CB}[1]{%
  \ifdim #1 pt > 0.849 pt \cellcolor{auc5}\color[RGB]{8,80,65}\textbf{#1}%
  \else\ifdim #1 pt > 0.749 pt \cellcolor{auc4}\color[RGB]{8,80,65}\textbf{#1}%
  \else\ifdim #1 pt > 0.649 pt \cellcolor{auc3}\color[RGB]{8,80,65}\textbf{#1}%
  \else\ifdim #1 pt > 0.549 pt \cellcolor{auc2}\color[RGB]{8,80,65}\textbf{#1}%
  \else \cellcolor{auc1}\color[RGB]{8,80,65}\textbf{#1}%
  \fi\fi\fi\fi}

\definecolor{corr4}{RGB}{142,206,179}  %
\definecolor{corr3}{RGB}{195,228,195}  %
\definecolor{corr2}{RGB}{197,218,238}  %
\definecolor{corr1}{RGB}{241,239,234}  %
\definecolor{corr0}{RGB}{247,193,193}  %

\newcommand{\R}[1]{%
  \ifdim #1 pt > 0.699 pt \cellcolor{corr4}\color[RGB]{8,80,58}#1%
  \else\ifdim #1 pt > 0.499 pt \cellcolor{corr3}\color[RGB]{8,80,65}#1%
  \else\ifdim #1 pt > 0.299 pt \cellcolor{corr2}\color[RGB]{4,44,83}#1%
  \else\ifdim #1 pt > -0.1 pt \cellcolor{corr1}\color[RGB]{95,94,90}#1%
  \else \cellcolor{corr0}\color[RGB]{121,31,31}#1%
  \fi\fi\fi\fi}

\newcommand{\RB}[1]{%
  \ifdim #1 pt > 0.699 pt \cellcolor{corr4}\color[RGB]{8,80,58}\textbf{#1}%
  \else\ifdim #1 pt > 0.499 pt \cellcolor{corr3}\color[RGB]{8,80,65}\textbf{#1}%
  \else\ifdim #1 pt > 0.299 pt \cellcolor{corr2}\color[RGB]{4,44,83}\textbf{#1}%
  \else\ifdim #1 pt > -0.1 pt \cellcolor{corr1}\color[RGB]{95,94,90}\textbf{#1}%
  \else \cellcolor{corr0}\color[RGB]{121,31,31}\textbf{#1}%
  \fi\fi\fi\fi}

\interspeechcameraready

\title{Comparator Loss: An Ordinal Contrastive Loss to Derive a Severity Score for Speech-based Health Monitoring}

\author[affiliation={1}]{Jacob J}{Webber}
\author[affiliation={1,2}]{Oliver}{Watts}
\author[affiliation={1}]{Lovisa}{Wihlborg}
\author[affiliation={2,3}]{Johnny}{Tam}
\author[affiliation={2,3}]{Christine}{Weaver}
\author[affiliation={2,3,4}]{Suvankar}{Pal}
\author[affiliation={2,3,4}]{Siddharthan}{Chandran}
\author[affiliation={1}]{Cassia}{Valentini-Botinhao}

\affiliation{}{SpeakUnique Ltd.}{UK}
\affiliation{Anne Rowling Regenerative Neurology Clinic}{University of Edinburgh (UoE)}{UK}
\affiliation{Institute of Adaptive and Neural Computation}{UoE}{UK}
\affiliation{}{UK Dementia Research Institute}{UK}
\email{jacob@speakunique.co.uk}

\keywords{Health monitoring, severity prediction, contrastive loss, representation learning}

\usepackage{comment}
\usepackage{soul}
\begin{document}

\maketitle

\begin{abstract}
Monitoring the progression of neurodegenerative disease (NDD) has important applications in planning treatment and evaluating new medications. Whereas much work has focused on discriminating patients from healthy controls, or predicting real-world health metrics, we propose a novel measure of disease progression: the severity score, derived from a model trained to minimize what we call the comparator loss. This loss ensures scores obey an ordering relation, based on diagnosis, clinical scores, or simply chronological order of recordings. The proposed comparator loss-based system has the potential to incorporate information from disparate health metrics, critical for making full use of small health-related datasets. We show that a model trained on lightly annotated data is capable of distinguishing between subjects with NDDs and healthy controls. Our score also correlates with annotations not observed in training, such as ALSFRS-R and those of speech and language therapists. 
\end{abstract}

\section{Introduction}
\label{sec:intro}

Neurodegenerative disorders are becoming more prevalent globally, posing a significant health challenge.
They include progressive and fatal diagnoses affecting cognition such as Alzheimer's disease, and neuromuscular disorders such as motor neuron disease (MND), the most common subtype of which is amyotrophic lateral sclerosis (ALS).
There is a growing need for diagnostic and monitoring methods that are both accurate and accessible \cite{ferrari2024global}.
Recent advances in speech technology and machine learning have highlighted the potential of using speech for detecting diseases and predicting their severity \cite{bowden2023systematic, de2020artificial}. 
Speech offers great potential as an objective digital biomarker for these conditions, being non-invasive and easily collected remotely without extensive expertise. 
State-of-the-art speech-based health monitoring methods focus on either categorising conditions (typically as healthy versus diagnosed) or predicting scores equivalent to those from existing medical tests. 
Classification tends to be the focus of most work in the area, epitomised by Challenge tasks such as \cite{bib:LuzHaiderEtAl20ADReSS}. For an overview of recent progress in predicting health status between healthy controls and patients with MND and Alzheimer's, see \cite{bowden2023systematic} and \cite{de2020artificial}.
Speech can also be used to monitor progression of a disease, in which case the task would entail regressing to a clinical score. 
These scores include, for ALS, ALSFRS-R \cite{cedarbaum1999alsfrs} and ECAS \cite{crockford2018specific}; and for dementia, ACE-III \cite{mathuranath2000brief}.

Although classifying speech as healthy or non-healthy might be simpler, this result alone offers limited insight and becomes particularly problematic if inaccurate.
Conversely, an automatic tool that replicates a clinician's score is only as effective as the score itself. Clinically annotated scores have been found unreliable and inconsistent across raters and centres. Authors in \cite{voustianiouk2008alsfrs} found that ALSFRS-R correlates poorly with another clinical score as the disease progresses. The work in \cite{bakers2022using} shows how different trial centres use the ALSFRS-R scale without perfect consistency.
In \cite{franchignoni2015further}, the authors discuss problems with the ALSFRS-R being aggregated from several items (like speech, stairs and food) with different metric properties.
Rather than learning annotated scores directly, the authors in \cite{Cummins2024} propose a model that predicts longitudinal shifts in depression from two speech samples. Even though the model is not learning to predict clinical scores, the shift directions used for training were obtained from clinically annotated data, the eight-item patient health questionnaire \cite{phq8}. 

Our proposed method follows a similar concept. Instead of predicting severity scores through regression to clinical scores, we aim to learn scores indirectly based on the order we believe governs the data.
To do that we introduce a new contrastive loss for neural network training, the comparator loss, aimed at operating on a continuous-valued score that can be used to indicate disease severity, progression, or simply diagnosis status. The comparator loss relies on the existence of an ordering system. An ordering system can be derived from clinical test scores, more simply from the time since diagnosis, or in its most simple form from the presence of a diagnosis.
Our loss is designed to impose an order on these scores based on the relationship between speech samples.
Specifically, it enforces that samples that are ranked higher, based on a particular ordering system, are also scored higher and vice versa. 

The proposed method offers several advantages over state-of-the-art classification techniques. 
The comparator loss enables the development of scoring models that can learn from different sources of potentially noisy information, extracting deeper insights from them.
It provides not only a more detailed picture than a discrete classification loss, but also facilitates fine-tuning across different tasks, which is crucial when dealing with small health-related datasets. 
Typical classifier-based models output a set of logits, meaning the number of classes is hard-coded into the model and its weights. 
Conversely, the comparator loss requires only the existence of an order, regardless of the number of classes. A model trained with this loss does not require `structural' changes to handle data with a different number of classes.

\section{Background}

\subsection{Category learning}

Categorical losses, such as cross-entropy, are well-established in the field of machine learning and commonly used in medical tasks. These make no assumption on the ordering of classes, meaning there is no inductive bias encoded in the model that categories are sequential. However, cross-entropy has also been applied to ordinal classes, such as the bits in binary encoding \cite{wavernn}. 
The cross-entropy loss requires the model to assign a probability to each class, with the condition that these sum to one. The loss is calculated as the difference between the predicted distribution and an ideal distribution, where the ideal distribution is taken to be a one-hot vector encoding the correct category. 
Authors in \cite{ordinal_category} extend the cross-entropy loss by incorporating ordering knowledge: rather than a one-hot target vector, ones are assigned to all categories up to and including the ground-truth category and zeros are assigned thereafter, with the loss computed via MSE or relative entropy.

\begin{figure}
    \centering %
    \includegraphics[width=0.99\linewidth]{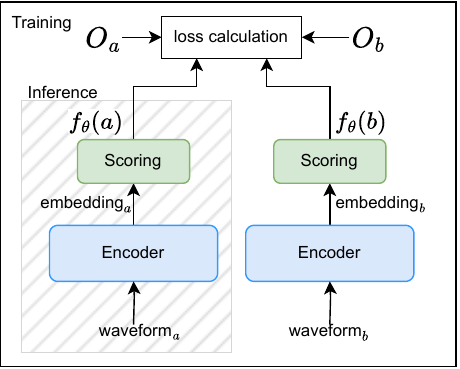}  %
    \vspace{-5pt}
    \caption{Model structure. During training the model learns to generate scores for pairs of speech waveforms which respect a known ordering $O$ (`comparator' loss). Hatched box: at inference time the model produces scores from unpaired waveforms.}
    \vspace{-10pt}
    \label{fig:model}
\end{figure}

\subsection{Representation learning and contrastive losses}

Representation learning is a suite of techniques used to learn representations -- typically a vector assigned to each sample -- for later downstream tasks. The relationship between different data samples can be used to guide representation learning. 
For instance, in contrastive representation learning \cite{weng2021contrastive}, similar samples are encouraged to be close in the embedding space, while dissimilar samples are encouraged to stay apart. 
An early example is the work described in \cite{contrastive} which introduces a contrastive loss for dimensionality reduction. Their proposed model is trained to learn a low-dimensional manifold from images. Pairs of images that are of similar categories are mapped to nearby points in that manifold. To prevent the representation space collapsing, a contrastive term is introduced, forcing image pairs from different categories apart.
Authors in \cite{Schroff_2015_CVPR} extend this work by proposing a triplet-based contrastive loss, whereby at each training step a training sample is compared to both a positive and negative sample (i.e.\ a sample within and outwith the category of the sample under consideration respectively). This was further extended into the so-called lifted structured loss \cite{lifted_structured_contrastive_all_pairs}, which compares all possible pairs in a batch for the purposes of calculating the contrastive loss.
Unsupervised learning methods use contrastive learning by relying on data augmentation for creating positive and negative sample pairs \cite{weng2021contrastive}.

While cross-entropy and MSE losses are commonly applied to speech-based health detection, some prior work has applied ordinal approaches to speech analysis.
A key example is the UTMOS model \cite{utmos} developed for synthetic speech evaluation. 
Unlike typical analysis tasks like disease detection and automatic speech recognition where performance can be measured using accuracy against a ground truth label, generative tasks like speech synthesis have historically relied on human evaluation. 
The most common type of evaluation is the Mean Opnion Score (MOS) test, where participants are asked to rate speech using a discrete 1--5 scale. 
The UTMOS model automates the evaluation process, and in doing so reduces the costs and potential for bias that come with using human evaluators \cite{huang2024mos}. 
Like the proposed comparator loss method, UTMOS employs an ordinal contrastive loss during training, where the difference between two ground truth scores is used to order samples, with a training loss used to ensure predicted score ordering agrees with ground truth. As with the comparator loss, the total minibatch loss is calculated by considering the sum of losses for every available pair of samples in the minibatch. 
The proposed method differs from UTMOS in several key respects. Beyond differences in application, input features, and training setup, the method we propose here does not rely on an additional MSE regression loss. This is significant as it allows the comparator loss to be trained on multiple ground truth signals that need not share the same scale, with comparisons performed per batch wherever a score exists for a given pair of samples.

\setlength{\tabcolsep}{5pt} %
\begin{table*}[t]
\centering
\caption{Datasets used. Classes: healthy control (HC), inaudible (INAUD), audible (AUD), motor neuron disease (MND), Parkinson’s disease (PD), amyotrophic lateral sclerosis (ALS). $^*$Only 70/248 speakers were annotated by SLPs (speech-language pathologists).}
\begin{tabular}{l l r r r r r}
\hline
& Dataset & Classes & Language & \#Speakers & Annotation (observed range) & \#Files \\ \hline
\multirow{2}{*}{\rotatebox[origin=c]{90}{Train}} 
& Proprietary (train)       & HC/INAUD/AUD & English (EN) & 367/356/388 & - & 371,359 \\
& Proprietary (dev)         & HC/INAUD/AUD & EN           & 44/40/37    & - & 41,288 \\ \hline
\multirow{7}{*}{\rotatebox[origin=c]{90}{Test\hspace{1pt}}}
& Proprietary                & HC/INAUD/AUD & EN          & 106/128/123 & - & 116,829 \\
& ARNSC \cite{ClinicData2025}& HC/MND/PD    & British EN  & 183/115/100 & ALSFRS-R speech (1-4) & 3,612 \\
& SAP \cite{sap}             & ALS/PD       & American EN & $^*$271/354 & SLP ratings$^*$ (1-7) & 244,115 \\
& DZNE \cite{dzne}           & HC/ALS       & German      & 51/72       & ALSFRS-R speech (1-4) & 185 \\
& VB2023 \cite{vb2023}       & HC/ALS       & Mandarin    & 61/39       & Dysarthria degree (2-4) & 12,751 \\
& EWA \cite{EWA24}           & HC/PD        & Slovak      & 95/95       & -              & 950 \\
& Neurovoz \cite{NeuroVoz24} & HC/PD        & Spanish     & 55/52       & -             &  1,695\\ \hline
\end{tabular}
\label{table:datasets}
\end{table*}

\section{Method}

\subsection{Model}

Fig.\,\ref{fig:model} illustrates the general structure of the proposed comparator model. The model is composed of an encoder that outputs an embedding given a speech waveform, and a scoring network that takes this embedding and converts it into a scalar value $f_\theta(.)$. 
We refer to this value as the `score' as it can be used as a proxy for a severity score. %
During training the model learns to generate a score for an input waveform by minimising a `comparator' loss. This loss is updated based on pairs of predicted scores and the `orders' $O$ associated with the inputs that produced them. 
While training requires pairs of stimuli, during inference (shown by the hatched box in Fig.\,\ref{fig:model}) the model produces scores from a single input waveform. 

\begin{figure}[t]
    \centering
    \vspace{-10pt}
    \includegraphics[trim={0 0 50 0},clip, width=0.9\linewidth]{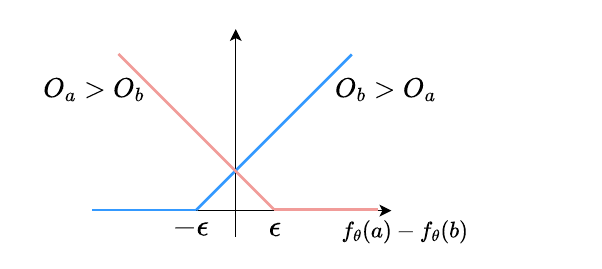}
    \vspace{-10pt}
    \caption{The comparator loss function. $O_{a}$, $O_{b}$ is the ground-truth ordinal value associated with sample $a$ and $b$. $f_{\theta}$ is the model as applied to the sample. $\epsilon$ is a hyperparameter. The curves show the loss in each ordinal case for a pair of samples. The loss penalises situations where the orders of the ordinal values and the orders of the model outputs are not in agreement.}
    \label{fig:loss}
    \vspace{-10pt}
\end{figure}

\subsection{Proposed loss function}
\label{sec:prop_loss_funk}

Fig.\,\ref{fig:loss} illustrates the comparator loss curves given the relative difference between predicted scores $f_\theta(a) -f_\theta(b)$.
The comparator loss is designed to ensure that predicted scores reflect the ordering of the elements, so that an element that is ranked higher has a greater score. The proposed loss function is similar to a traditional contrastive loss \cite{contrastive}, with two key differences. The first is that instead of predicting a vector and using cosine differences, it uses a model that outputs a single scalar value, assigning a single score to samples. The second is that rather than assigning similarity or difference based on whether a pair of samples are from similar or different categories, it instead depends on an assigned \emph{order} to the categories, penalizing results where for a pair of samples, a higher score is given to the sample of a lower order of category.

For each pair of samples $(a, b)$, there are three possible scenarios: $O_a>O_b$, $O_a = O_b$ or $O_b > O_a$, where the in/equalities are determined solely by the order of the samples' classes. By swapping the ordering of the samples in the pair we can ensure that $O_b\geq O_a$, reducing the number of cases to two. For pairs of samples in the same class, we define the loss as zero. Otherwise, for when $O_b > O_a$, the loss is defined as:

\begin{equation}
    J = \max((f_\theta(a) -f_\theta(b)) + \epsilon, 0)
\end{equation}
\noindent where $\epsilon$ is a hyper-parameter designed to stop rewarding the model once the scores are correctly ordered beyond a certain threshold. 
We can see in Fig.\,\ref{fig:loss} that when $O(b) > O(a)$ (blue curve) the loss is zero when $f_\theta(b) - f_\theta(a) > \epsilon$.

The scores output by the model are not anchored to any pre-defined rating scale. Their absolute magnitude will depend on factors such as the margin hyperparameter $\epsilon$ and the network weight initialisation scheme. While we demonstrate below that the scores are already meaningful in that they correlate with accepted clinical scales, it may be desirable to impose an interpretation on their absolute values. One option would be to constrain the model output to a pre-specified range by modifying the output layer accordingly. More promisingly, we are investigating post-hoc calibration of the score by transforming it to match the distribution of a large, representative speech sample. Decision thresholds could then be expressed as standard deviations from zero, where zero represents the population mean, providing an interpretable and clinically communicable scale.

\section{Evaluation}

\subsection{Data}

Datasets used for training and testing are described in Table \ref{table:datasets}.

\subsubsection{Training data}\label{section:training_data}

We used a proprietary dataset collected in the course of a commercial voicebanking operation to train and validate the model. People who expect to lose their voice due to medical conditions, and whose speech may already be impaired, can register and record a speech sample which is used as the basis for a personalised synthetic voice for use with communication aids. Speech is elicited using a prompt script 
and recorded remotely via a web platform; users are advised to record in a quiet, non-reverberant environment, but recording conditions are otherwise uncontrolled and variable. On average, each speaker contributes approximately $40$ minutes of speech.

Users with a speech-affecting condition select a product tier based on whether they judge their speech to be already affected at the time of registration. Users with no known speech-affecting condition may also register to explore the service, which is often done by healthcare professionals demonstrating the service to patients. The database is supplemented with purpose-made recordings of healthy volunteers collected under more controlled conditions.

Using the product tier selected at registration, we lightly annotate speakers as belonging to one of three groups, those with: (1) no known speech-affecting condition (which we refer to as HC, healthy control); (2) a condition expected to affect speech in future, but with no self-reported signs at time of recording (which we refer to as INAUD that stands for inaudible); and (3) a condition already affecting speech at time of recording (which we refer to as AUD as audible). Permission was obtained to use speech recordings for this experiment with self-reported sex and age. While users may provide additional health information at registration, consent to use that information for current purposes was not obtained for this cohort, and no further annotation was used. A more recent cohort from the same service, for which such consent was obtained, suggests that just over half of registered users have MND, approximately a quarter have Parkinson's or parkinsonian conditions, and the remainder have other conditions including neurological disorders, head and neck cancer, and other causes of voice impairment.

From a pool of speakers whose data was gathered in this way, a balanced subset was sampled using Coarsened Exact Matching (CEM) \cite{iacus2012causal} to ensure the three groups (HC, INAUD, AUD) were comparable with respect to sex and age. 
It should be noted that matching on age and sex has an unintended consequence for the HC group: because purpose-made recordings of healthy volunteers were needed to achieve balance, the HC group contains a higher proportion of controlled-condition recordings than the INAUD and AUD groups. This introduces a likely confound whereby models may partially learn to discriminate health status on the basis of recording condition rather than speech characteristics alone. The use by statistical models of recording environment as a proxy for health status is a known problem in speech-based health monitoring; for example \cite{schu2023using} and \cite{liu2024clever} show that for several widely-used benchmark datasets, health status can be classified at well above chance from non-speech segments of recordings alone due to case and control participants being recorded in a systematically different way. We are aware of this effect in our dataset but note that this confound is specific to the HC group: since INAUD and AUD speakers are drawn from the same voicebanking service under comparable recording conditions, and CEM sampling introduces no systematic bias between these two groups, discrimination between them is unlikely to be affected. We discuss the potential impact on results involving the HC group in Section \ref{sec:results}.

The final dataset contains recordings from 1,589 speakers divided (with stratification in terms of the three classes) into training (70\%, totalling 1,111 speakers), development (10\%, 121 speakers), and test (20\%, 357 speakers) sets.

\subsubsection{Test data}

Test datasets used for evaluation are listed in the lower portion of Table \ref{table:datasets}. For some of the datasets we used a selection of data available in order to focus on the conditions detailed in the \textit{classes} column and on a subset of speakers where annotation was available. They were selected to evaluate generalisation across a deliberately heterogeneous range of conditions, languages, and annotation schemes. Beyond the held-out proprietary test set, which matches the training distribution, the remaining datasets span six language varieties (British and American English, German, Mandarin, Slovak, and Spanish), two target conditions (MND/ALS and PD), and five distinct annotation schemes -- including the ALSFRS-R speech subscale \cite{cedarbaum1999alsfrs}, dysarthria degree \cite{vb2023}, and speech and language pathologists' ratings \cite{sap}. No annotations from any of these datasets were used during training. This heterogeneity is a deliberate design choice: a clinically useful severity score should generalise across languages and recording conditions, and should correlate with annotation schemes it has never been directly trained to predict.
This stands in contrast to much prior work in speech-based health monitoring, where results are reported via cross-validation or train/test splits of a single dataset. Such evaluations risk overstating performance by conflating in-distribution generalisation with real-world robustness, and may not reflect the shifts in language, recording condition, and clinical context that any deployed system would inevitably encounter.

\subsection{Training Setup}
All models were trained for 100 epochs with a batch size of 64 using the Adam optimiser. Model checkpoints were saved only when validation accuracy exceeded all previously recorded values. Validation accuracy was defined as the accuracy of discriminating between INAUD and AUD speakers -- the most challenging pair of classes -- using a support vector classifier trained on the output severity scores. Hyperparameter selection for both the choice of encoder and scoring model was performed via a series of sweeps using this validation metric.

\subsection{Encoder}
We used embeddings from pre-trained encoders, motivated by their strong performance on health diagnosis tasks \cite{kitchen_sink26} and by prior observations that frozen pre-trained encoders outperform encoders trained jointly from scratch with the scoring network on this task. Only the scoring network was fine-tuned during training. We tested all combinations of one to three of the following encoders: MMS \cite{mms}, HuBERT \cite{hubert}, Whisper \cite{whisper}, TitaNet \cite{TitaNet}, UniSpeech \cite{unispeech}, x-vector \cite{tddxvector}, Wav2Vec2+Conformer \cite{wav2vec2conformer}, and CRDNN+CTC \cite{speechbrainenc-decasr}. When multiple encoders were used, their embeddings were concatenated to form the input representation. Although approaches such as LoRA-based adaptation may yield further improvements, investigating this is left to future work as the focus of this paper is the loss function.

\subsection{Scoring Network}
The scoring model uses a simple feed-forward architecture consisting of $N$ blocks. Each block consists of a linear layer, followed by layer normalisation, dropout and LeakyReLU activation. The blocks are followed by a final linear layer that maps the output representation into one value (or into a value for each class in the case of the Cross-Entropy baseline). We tested $N$ varying from 2 to 4 of varying sizes $M$ (128, 256, 512, 1024 and 2048), keeping the size of the layer of the last block always fixed as 64. 

\subsection{Proposed Loss}
For the proposed comparator loss, scores were generated for every example in a batch and the loss computed for all possible pairs within the batch. The sum of pairwise loss terms was used to update the model. The margin hyperparameter $\epsilon$ was set to the value used in \cite{contrastive}. The computational cost of the all-pairs approach scales quadratically with batch size; in practice we found that a batch size of 64 provided a good trade-off between loss coverage and training efficiency. %

\subsection{Baseline losses}
Three baselines were constructed using alternative loss functions. The \textit{MSE} baseline minimises mean squared error between the predicted output and a numeric class label (0: HC, 1: INAUD, 2: AUD). The \textit{Contrastive} baseline uses the original contrastive loss \cite{contrastive}, trained to produce maximally separated scores for distinct classes and equal scores for identical classes. The \textit{Cross-Entropy} baseline uses cross-entropy loss to classify speakers into the three groups. For the Contrastive and MSE baselines, the scalar model output is used directly as the severity score. 
For the Cross-Entropy baseline, the severity score is derived as a weighted average of canonical class values 0, 1, and 2 assigned to HC, INAUD, and AUD respectively, with predicted class probabilities used as the weights. We note that in contrast to the underlying classification model, this conversion injects knowledge of how classes should be ordered with regards to severity.

\setlength{\tabcolsep}{4pt} %
\begin{table}[t]
\centering
\caption{Best configuration for encoder, scoring model ($N$, $M$), dropout and learning rate scheduler. $N$ is the number of blocks of the scoring model and $M$ the size of the first $N$-1 blocks.} %
\begin{tabular}{l r r r r r}
\hline
Loss & Encoder & $N$ & $M$ & Dropout & Scheduler \\ \hline
Proposed %
 & HuBERT & 3 & 512 & 0.5 & Cyclic \\ 

Cross-entropy %
 & HuBERT & 3 & 512 & 0.5 & Cyclic \\ 

Contrastive & HuBERT & 3 & 512 & 0.4 & - \\ 

MSE & Whisper & 3 & 1024 & 0.5 & Cyclic \\

\end{tabular}
\label{table:best_sweeps}
\end{table}

\begin{table*}[t]
\centering
\caption{\textit{AUC for several binary tasks of different datasets. Colours reflect: \colorbox{auc5}{\color[RGB]{8,80,65}${\geq}.85$}, \colorbox{auc4}{\color[RGB]{8,80,65}${\geq}.75$}, \colorbox{auc3}{\color[RGB]{4,44,83}${\geq}.65$}, \colorbox{auc2}{\color[RGB]{65,36,2}${\geq}.55$}, \colorbox{auc1}{\color[RGB]{80,19,19}${<}.55$}.}}
\begin{tabular}{l ccc cc c c c c}
\toprule
& \multicolumn{3}{c}{\textbf{Proprietary}} & \multicolumn{2}{c}{\textbf{ARNSC}} & \textbf{DZNE} & \textbf{VB2023} & \textbf{EWA} & \textbf{Neurovoz} \\
\cmidrule(lr){2-4} \cmidrule(lr){5-6}
\textbf{Loss} & \textit{HC/INAUD} & \textit{HC/AUD} & \textit{INAUD/AUD} & \textit{HC/MND} & \textit{HC/PD} & \textit{HC/ALS} & \textit{HC/ALS} & \textit{HC/PD} & \textit{HC/PD} \\
\midrule
Proposed      & \CB{0.97} & \CB{0.998} & \CB{0.93} & \C{0.84}  & \CB{0.73} & \C{0.74}  & \C{0.74}  & \CB{0.64} & \CB{0.84} \\
Cross-entropy & \C{0.97}  & \C{0.998}  & \CB{0.93} & \C{0.85}  & \C{0.72}  & \CB{0.77} & \C{0.72}  & \C{0.63}  & \C{0.65} \\
Contrastive   & \C{0.55}  & \C{0.947}  & \C{0.88}  & \C{0.79}  & \C{0.59}  & \C{0.37}  & \C{0.34}  & \C{0.52}  & \C{0.66} \\
MSE           & \C{0.96}  & \C{0.996}  & \C{0.92}  & \CB{0.87} & \C{0.65}  & \C{0.73}  & \CB{0.88} & \C{0.55}  & \C{0.80} \\
\bottomrule
\end{tabular}
\label{tab:auc}
\end{table*}

\begin{table*}[t]
\centering
\caption{\textit{Spearman correlation between speaker-level predicted scores and clinical annotations.
ARNSC and DZNE datasets use the ALSFRS-R speech subscale that decreases with impairment; correlation signs are flipped for display so that
positive values indicate correct alignment with impairment severity.
Colours reflect:
\colorbox{corr4}{\color[RGB]{8,80,58}${\geq}0.70$},
\colorbox{corr3}{\color[RGB]{8,80,65}${\geq}0.50$},
\colorbox{corr1}{\color[RGB]{95,94,90}${\approx}0$}.
$^{\dagger}$Signs flipped for display.}}
\begin{tabular}{l cccc c c}
\toprule
& \textbf{ARNSC} & \multicolumn{3}{c}{\textbf{SAP}} & \textbf{DZNE} & \textbf{VB2023} \\
\cmidrule(lr){3-5}
\textbf{Loss} & \textit{ALSFRS-R sp.}$^{\dagger}$ & \textit{Natural.\ (ALS)} & \textit{Natural.\ (PD)} & \textit{Imprecise cons.} & \textit{ALSFRS-R sp.}$^{\dagger}$ & \textit{Dysarthria deg.} \\
\midrule
Proposed      & \RB{.73} & \RB{.76} & \RB{.64} & \RB{.78} & \RB{.75} & \R{.65} \\
Cross-entropy & \R{.71}  & \R{.75}  & \RB{.64} & \R{.76}  & \R{.73}  & \R{.66} \\
Contrastive   & \R{.72}  & \R{.67}  & \R{.53}  & \R{.66}  & \R{-.01}  & \R{-.07} \\
MSE           & \R{.71}  & \RB{.76} & \R{.58}  & \R{.71}  & \R{.69}  & \RB{.74} \\
\bottomrule
\end{tabular}
\label{tab:spearman}
\end{table*}

The best-performing configuration for each loss function is reported in Table\ \ref{table:best_sweeps}. Although the MMS encoder achieved competitive performance on the proprietary test set, it was excluded from the final system on the grounds that its license does not permit commercial use. In all cases, a non-MMS configuration performed comparably on the proprietary test set, and these configurations are reported in Table \ref{table:best_sweeps}.

\subsection{Metrics}
\label{sec:metrics}

We evaluate our systems based on the following metrics:
\begin{itemize}[left=0pt]
    \item \textbf{Area under the receiver operating characteristic curve (AUC)} on the task of classifying between speech from healthy controls and people with a diagnosis of either MND or PD. We expect that a true severity score should be able to categorise samples into those classes with a high degree of accuracy. %
    \item \textbf{Correlation}: the severity score should correlate with existing clinical scores for tracking disease. Correlation is measured using Spearman rank correlation.

\end{itemize}

All metrics presented are calculated at a speaker level. Speaker level severity scores were obtained by averaging the severity score across all recordings of a speaker. Speaker level clinical annotation were obtained by taking the median of the clinical annotation of all recordings of a speaker (when that was available).

\section{Results}\label{sec:results}

Table \ref{tab:auc} presents AUC values for binary classification across all test datasets, and Table \ref{tab:spearman} presents Spearman rank correlations between predicted scores and clinical annotations not observed during training. The two tables are complementary, and do not exactly overlap in their coverage of the datasets; SAP contains no healthy control speakers and therefore does not have associated classification results, while EWA and Neurovoz provide no speech-relevant rating scales and therefore do not appear in the correlation table.

\begin{figure*}[h!]
    \noindent\hspace{-15pt}%
    \begin{subfigure}[t]{0.35\linewidth}
        \includegraphics[width=\linewidth, trim=0 30pt 0 0, clip]{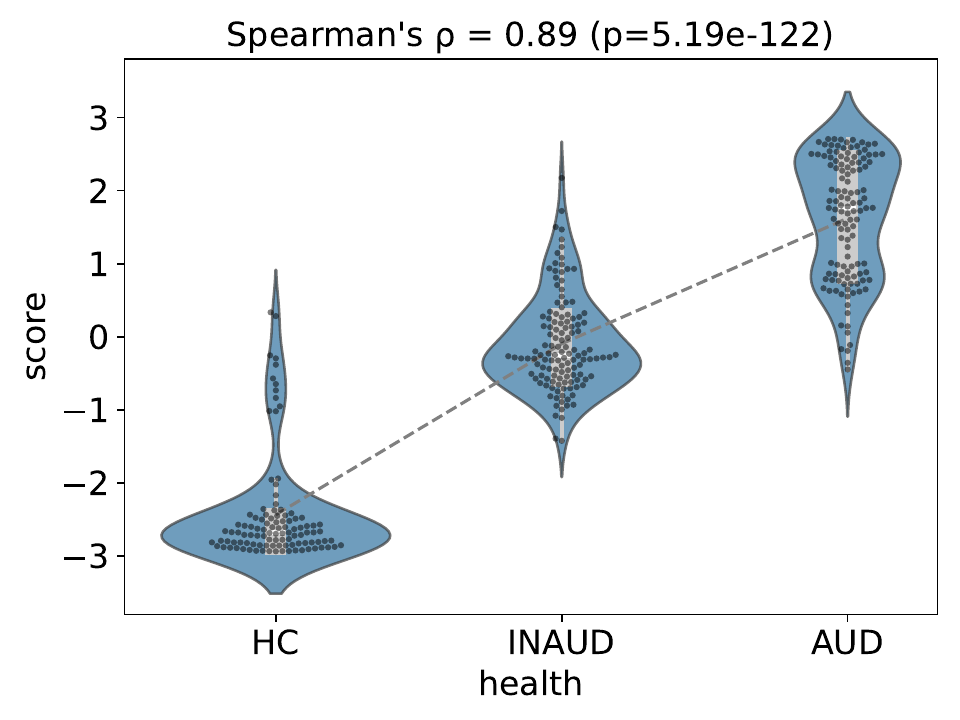}
        \caption*{\hspace{20pt}Self-reported speech disturbance (Proprietary)}
    \end{subfigure}%
    \begin{subfigure}[t]{0.35\linewidth}
        \includegraphics[width=\linewidth, trim=0 30pt 0 0, clip]{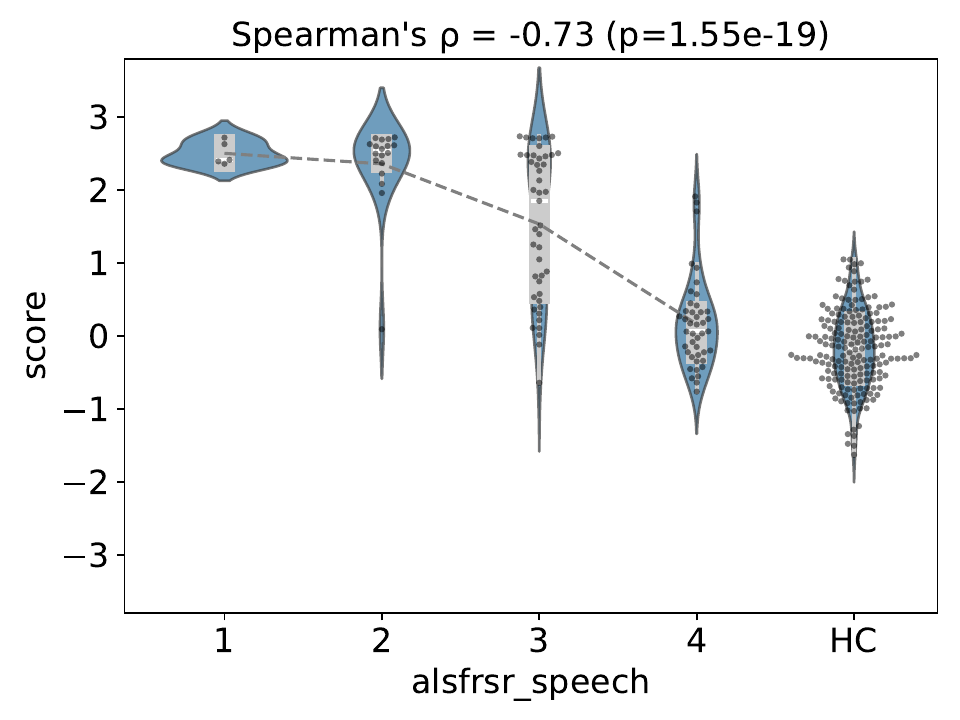}
        \caption*{\hspace{20pt}ALSFRS-R speech (ARNSC)}
    \end{subfigure}%
    \begin{subfigure}[t]{0.35\linewidth}
        \includegraphics[width=\linewidth, trim=0 30pt 0 0, clip]{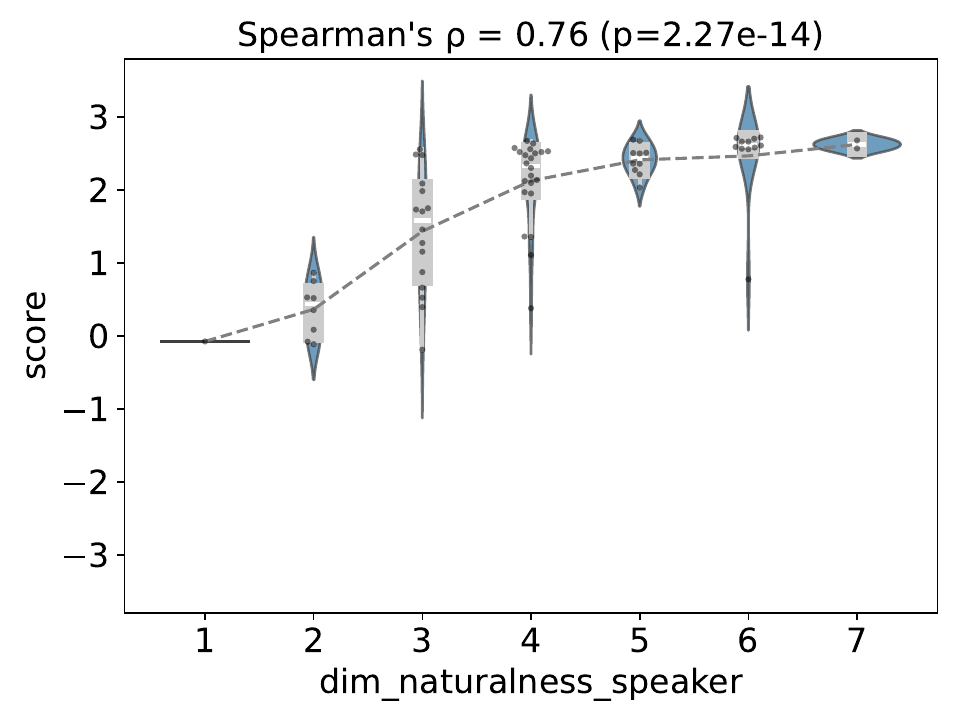}
        \caption*{\hspace{20pt}Naturalness ALS (SAP)}
    \end{subfigure}%
    \vspace{5pt}
    \par\noindent\hspace{-15pt}%
    \begin{subfigure}[t]{0.35\linewidth}
        \includegraphics[width=\linewidth, trim=0 30pt 0 0, clip]{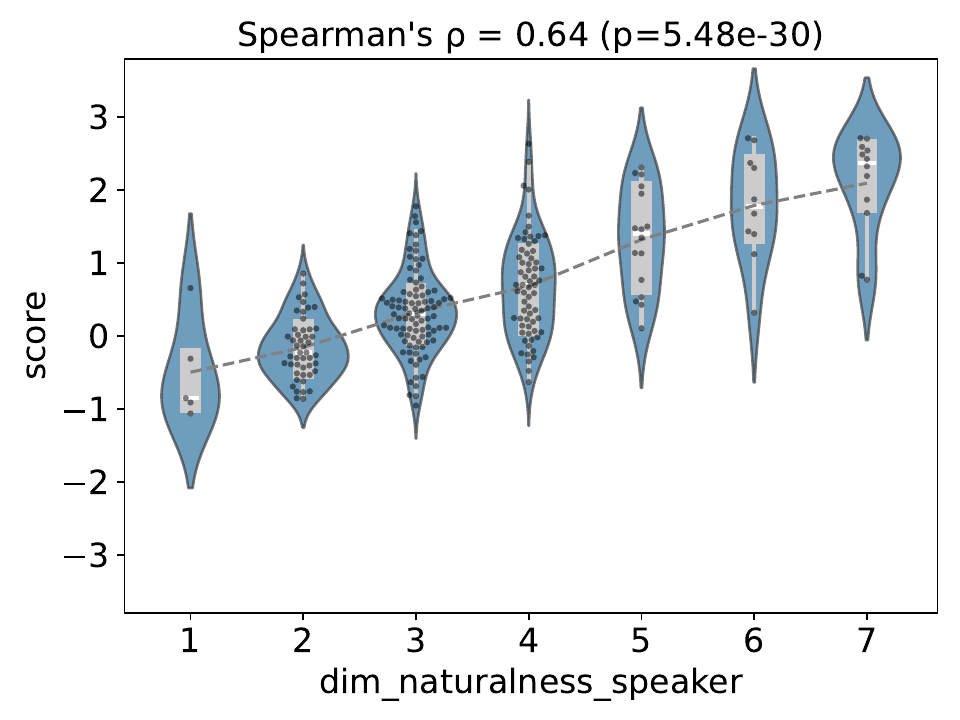}
        \caption*{\hspace{20pt}Naturalness PD (SAP)}
    \end{subfigure}%
    \begin{subfigure}[t]{0.35\linewidth}
        \includegraphics[width=\linewidth, trim=0 30pt 0 0, clip]{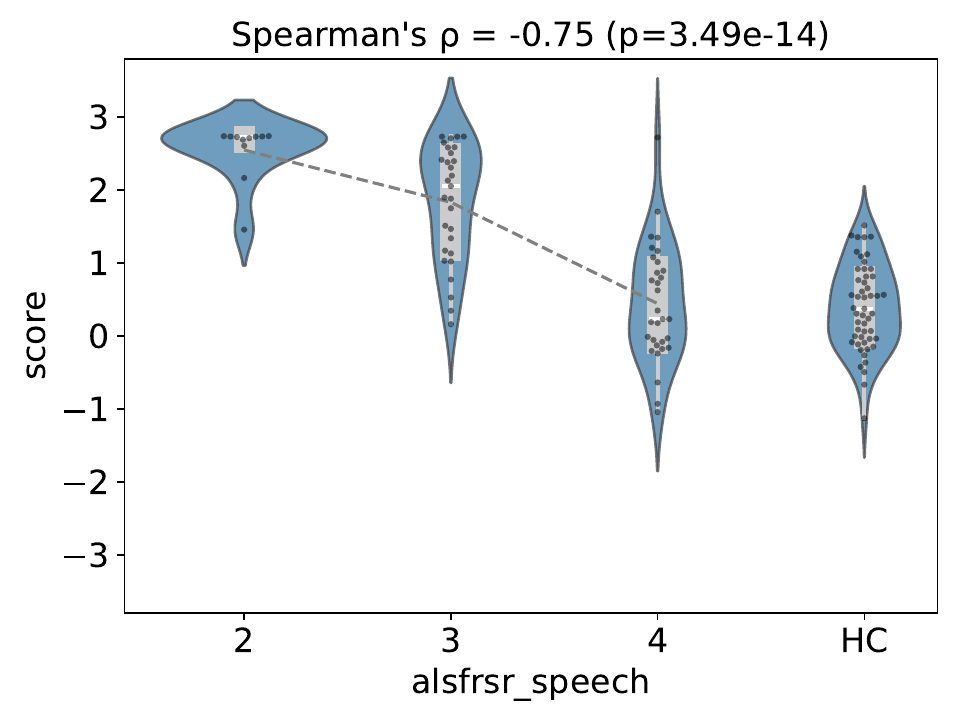}
        \caption*{\hspace{20pt}ALSFRS-R speech (DZNE)}
    \end{subfigure}%
    \begin{subfigure}[t]{0.35\linewidth}
        \includegraphics[width=\linewidth, trim=0 30pt 0 0, clip]{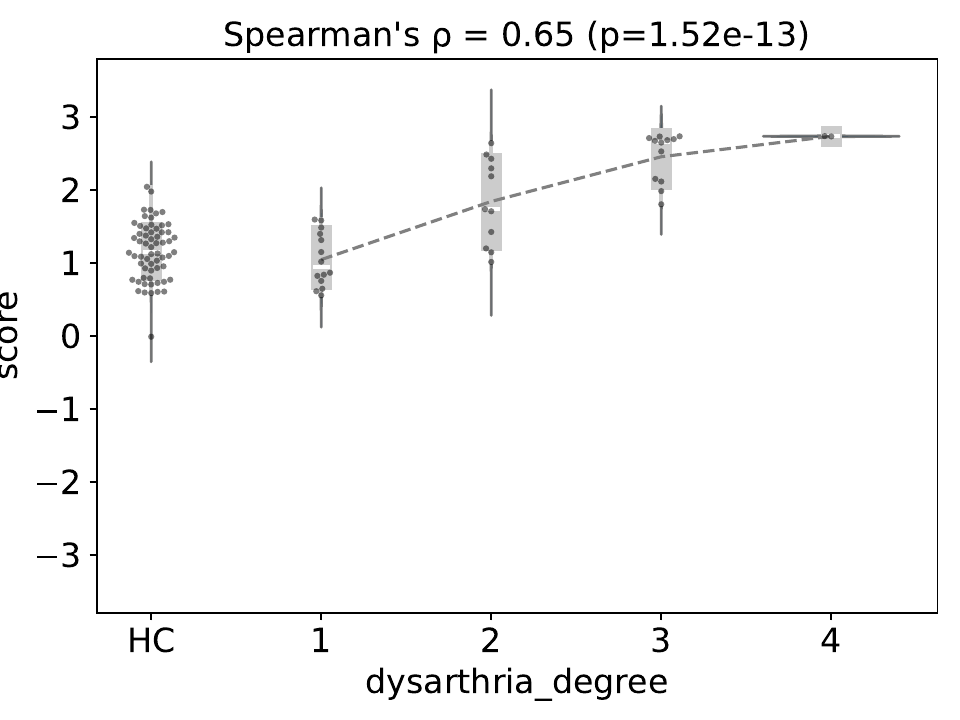}
        \caption*{\hspace{20pt}Dysarthria degree (VB2023)}
    \end{subfigure}%
    \caption{Violin plot of severity scores from best performing model (proposed loss) on various annotations. Datasets in parenthesis. ALSFRS-R speech (1–4): 4=normal, 3=detectable disturbance, 2=intelligible with repetition, 1=combined vocal/nonvocal communication. Naturalness and imprecise consonants (1–7): 1= normal, 7=severe deviation; low naturalness reflects unusual prosodic/rhythmic characteristics, imprecise consonants reflects slurring and distortion. Dysarthria degree (1–4): 1=fluent, 2=disfluent prosody, 3=mild dysarthria with high intelligibility, 4=dysarthria with low intelligibility.}
    \label{fig:violins}
    \vspace{-10pt}
\end{figure*}

\subsection{Classification}

The proposed loss is competitive with or outperforms all baselines on the majority of classification tasks. It achieves the highest AUC on HC/PD discrimination on Neurovoz (0.84 vs.\ 0.65 for cross-entropy), and performs comparably to cross-entropy and MSE on the proprietary and ARNSC datasets. The contrastive baseline is a notable exception: while it performs well on the proprietary INAUD/AUD task, it fails on both the in-domain HC/INAUD task and on out-of-domain datasets, falling near or below chance on DZNE and VB2023. We suspect both the performance on proprietary dataset tasks and failure on other datasets is due to confounding effect: because the contrastive loss imposes no directional constraint on the ordering of classes, the model may learn to place HC between INAUD and AUD rather than at one end of the severity continuum. This is plausible given that the three classes in the proprietary data differ not only in health status but also potentially in acoustic environment (see Section \ref{section:training_data}). Similar systematic differences in recording environment between healthy controls and patients may arise in the DZNE and VB2023 datasets, which could lead the undirected contrastive loss to produce an inverted or collapsed score ordering. Further investigation is needed to test this explanation.

\subsection{Correlation with Expert Annotations}
A key motivation for the comparator loss is its ability to learn a severity score that can track heterogeneous clinical annotations even when they are not seen during training. Table \ref{tab:spearman} demonstrates this across a diverse range of annotation: the ALSFRS-R speech subscale (available for both ARNSC and DZNE datasets), \textit{naturalness} and \textit{imprecise consonants} 
(low naturalness reflects unusual prosodic/rhythmic characteristics and imprecise consonants reflects slurring and distortion)
ratings from speech and language pathologists (SAP dataset), and a dysarthria degree scale (VB2023 dataset) -- spanning multiple languages, recording conditions, and rating methodologies.

The proposed loss achieves the highest or joint-highest correlation in five of the six reported conditions, with Spearman $\rho$ ranging from 0.64 to 0.78. 

The contrastive baseline again performs poorly, collapsing to near-zero correlation on DZNE and VB2023, consistent with the score inversion described above. MSE performs comparably to the proposed loss on several tasks but shows greater variability, falling notably behind on SAP naturalness for PD speakers. The proposed loss and cross-entropy are generally close, though the proposed loss shows a consistent small advantage on all tasks shown in Table \ref{tab:spearman}, where it receives highest or joint-highest correlation in five out of the six tasks.

All results across all four loss functions show that tasks involving PD speakers are consistently harder than those involving MND. While the task may be inherently harder, we also observe that MND is more richly represented than PD in the proprietary training data, reflecting the demographics of voicebanking users. Models trained on this data should therefore be expected to better characterise MND-related dysarthria.

Figure \ref{fig:violins} shows violin plots of scores generated by the best performing model using the proposed loss on the different levels of several annotations.
Each violin represents the distribution of predicted severity scores in that given level. The dashed line connects the median scores across categories -- a consistently rising or falling trend indicates good alignment with the annotation scale. Individual speaker-level data points are overlaid as dots.
It confirms that the high correlations achieved by the comparator loss based model reflect consistent monotonic trends across annotation levels, with median scores rising reliably as impairment severity increases.

We note that HC speakers occupy different score regions across datasets: most negative in the Proprietary dataset, higher in ARNSC and DZNE, and highest in VB2023.
The consistent directionality (HC correctly distinguished from impaired speakers within each dataset) is the meaningful result, while the value of the score of HC shifting across datasets reflects the model's sensitivity to acoustic differences that were not controlled for during training. This suggests that some form of score normalisation or domain adaptation is needed for deployment.

\section{Conclusions}

We have proposed the comparator loss, a novel ordinal contrastive loss for deriving a continuous severity score from speech. Models trained with this loss produce scores that correlate significantly with a range of clinical annotations -- including ALSFRS-R, speech-language pathologist ratings, and dysarthria degree scales -- none of which were observed during training, and across multiple languages and recording conditions. The proposed loss compares favourably with strong baselines, and we have established benchmark results across seven datasets spanning two target medical conditions, providing a reference point for future work in this area.

A key strength of the approach is its flexibility with respect to supervision signal. The comparator loss requires only the existence of an ordering relation between pairs of samples, and is indifferent to the scale, range, or provenance of the underlying annotation. This makes it straightforward to combine multiple heterogeneous and potentially weak annotation sources within a single training framework -- for example, combining clinical scores with the chronological order of longitudinal recordings from the same patient, or with coarse self-reported labels. We expect that exploiting this property more fully, by increasing the diversity and scale of training annotations, will yield further improvements. Confirming the ability of the score to track disease progression in longitudinal data remains another important direction for future work.

\bibliographystyle{IEEEtran}
\bibliography{mybib}

\end{document}